\documentclass{appolb}
\usepackage{graphicx}

\newcommand{\gtrsim}{\lower2pt\hbox{${\buildrel > \over \sim}$}}
\newcommand{\lesssim}{\lower2pt\hbox{${\buildrel < \over \sim}$}}

\begin{document}
\title{Multiwavelength and Multimessenger Observations of Blazars and Theoretical Modeling: Blazars as Astrophysical Neutrino Sources
\thanks{Presented at the XXVIII Cracow EPIPHANY Conference on Recent Advances in Astroparticle Physics, 10 -- 14 January 2022}%
}
\author{Markus B\"ottcher
\address{Centre for Space Research, North-West University, Potchefstroom, South Africa}
\\[3mm]
{Matthew Fu, Timothy Govenor, Quentin King, Parisa Roustazadeh\footnote{Also: Columbus State Community College, Columbus, OH, USA}
\address{Bishop Watterson High School, Columbus, OH, USA}
}
}
\maketitle
\begin{abstract}
This contribution reviews recent advances in the possible identification of blazars as potential sources of at least some of the very-high-energy 
neutrinos detected by the IceCube neutrino detector at the South Pole. The basic physical requirements for neutrino production and physics 
constraints that may be drawn from neutrino - blazar associations are reviewed. Several individual cases of possible associations will be 
discussed in more detail. It is emphasized that due to $\gamma\gamma$ opacity constraints in efficiently neutrino-producing blazars, an
association between X-ray --- soft $\gamma$-ray activity and very-high-energy neutrino production is more naturally expected than a
connection between neutrino and high-energy/very-high-energy $\gamma$-ray activity. 
\end{abstract}

\section{\label{sec:intro}Introduction}

Since the first announcement of the detection of very-high-energy neutrinos of astrophysical origin by the IceCube neutrino detector at the 
South Pole \cite{IceCube2013a,IceCube2013b}, the identification of the sources of these neutrinos has been one of the highest-priority 
quests in modern astrophysics. Various high-energy sources, including supernovae and their remnants, tidal disruption events, and active
galactic nuclei (AGN) have been suggested as possible sources of very-high-energy neutrinos (see \cite{HK22} and \cite{MS22} for recent 
reviews). Among the most promising candidates are blazars, a class of jet-dominated 
AGN with their relativistic jets pointing close to our line of sight, resulting in significant Doppler boosting of their emission. They are known
to be variable throughout the electromagnetic spectrum, from radio to high-energy (HE: $E > 100$~MeV) and in some cases even
very-high-energy (VHE: $E > 100$~GeV) $\gamma$-rays, on a variety of time scales, from months to years down to minutes. Blazars
consist of flat-spectrum radio quasars (FSRQs) and BL Lac objects, depending on the presence or absence of bright optical/UV emission
lines, respectively. Their broad-band spectral energy distributions are dominated by two broad, non-thermal emission components:
The low-frequency component (radio through optical/UV/X-rays) is likely due to synchrotron emission by relativistic electrons in the
jet, while the high-frequency (X-rays to $\gamma$-rays) component may be either due to inverse Compton scattering of low-energy
target photons by the same relativistic electrons, or due to hadronic processes (for a recent review see, e.g., \cite{Boettcher19}). 
Depending on the location of the peak of the synchrotron component, blazars are further sub-divided into low-synchrotron-peaked
(LSP; $\nu_{\rm sy} < 10^{14}$~Hz), intermediate-synchrotron-peaked (ISP; $10^{14}$~Hz~$< \nu_{\rm sy} < 10^{15}$~Hz)  
and high-synchrotron-peaked (HSP; $\nu_{\rm sy} > 10^{15}$~Hz) blazars. All known FSRQs belong to the LSP class, while BL Lacs
display synchrotron peaks in all ranges. 

While it has been demonstrated that blazars can not be responsible for the entire IceCube astrophysical neutrino excess (e.g., \cite{Murase18}), 
the number of plausible spatial and temporal associations between blazars in multi-wavelength outbursts and IceCube neutrino alerts is steadily 
increasing (see, e.g., \cite{Garrappa19}, \cite{Franckowiak20} and \cite{Giommi20b} for comprehensive multi-wavelength studies of plausible 
associations). 

This review will provide an introduction to the physical conditions and energetics required for VHE neutrino production in blazar jets 
(\S \ref{sec:energetics}) and then discuss the results of detailed studies of individual cases of potential blazar-flare -- neutrino-alert
associations (\S \ref{sec:individual}). \S \ref{sec:summary} will finally provide a summary of lessons learned from these studies.

\section{\label{sec:energetics}General considerations: Physical conditions and energetics requirements}

The production of $> 100$~TeV neutrinos (which have a high probability of being of astrophysical origin) requires the acceleration of 
protons or nuclei to at least $\sim$~PeV energies. Neutrinos are a byproduct of the decay of charged pions and muons produced in 
hadronic interactions dominated by the excitation and subsequent decay of the $\Delta^+$ (1232~MeV) resonance (see, e.g., Chapter
3 of \cite{BHK12}). In about 1/3 of the interactions, the $\Delta^+$ decay results in the production of a charged pion: 

\begin{equation}
p + p \to p + \Delta^+ \to p + n + \pi^+ \;\;\;\;\;\; {\rm or} \;\;\;\;\; p + \gamma \to \Delta^+ \to n + \pi^+
\label{eq:Delta}
\end{equation}
followed by
\begin{eqnarray}
\pi^+ \to \mu^+ + \nu_{\mu} \cr
\mu^+ \to e^+ + \nu_{e} + \overline{\nu_{\mu}}
\label{eq:decay}
\end{eqnarray}

In the jets of AGN, the number of potential target protons for $pp$ interactions is expected to be many orders of magnitude smaller
than the number of available target photons for $p\gamma$ interactions. Almost all works studying neutrino production in blazars
therefore only consider the photo-pion production channel (see, however, e.g., \cite{Liu19} for an alternative, hadronuclear, beam-dump-like 
scenario). The excitation of the $\Delta^+$ resonance requires the corresponding center-of-momentum energy of the reaction, thus
constraining the product of the proton ($E'_p$) and target photon ($E'_t$) energies in the co-moving frame of the emission and neutrino
production region, which moves with relativistic speed along the jet:

\begin{equation}
s \sim E'_p \, E'_t \, = E_{\Delta^+}^2 = (1232 \, {\rm MeV})^2  
\label{spgamma}
\end{equation} 

The relativistic motion of the emission region along the jet closely aligned with our line of sight leads to relativistic Doppler boosting
of both electromagnetic and neutrino emissions by a Doppler factor $\delta \equiv 10 \, \delta_1$. In the photopion reactions and subsequent
pion and muon decay, each neutrino takes up on average about 5~\% of the proton's energy. Consequently, the production of an IceCube
neutrino with observed energy $E_{\nu} = 100 E_{14}$~TeV (which is produced with $E'_{\nu} = 10 \, \delta_1^{-1} \, E_{14}$~TeV in
the emission-region rest frame), requires the acceleration of protons to $E'_p \sim 200 \, E_{14} \, \delta_1^{-1}$~TeV. The Delta
resonance condition (\ref{spgamma}) then implies that target photons of energies $E'_t \sim 1.6 \, E_{14}^{-1} \, \delta_1$~keV
(i.e., X-rays) are needed. 

Different constraints on the power in relativistic protons in the jet can be derived, depending on whether the target photon field
originates within the jet, co-moving with the emission region (e.g., the primary electron-synchrotron radiation), or outside the
jet, stationary in the AGN rest frame. In the former case, the co-moving target photons are Doppler boosted into the observer's
frame to an energy of $E_t^{\rm obs} \sim 16 \, E_{14}^{-1} \, \delta_1^2 / (1 + z)$~keV. In addition to the frequency shift,
the hard X-ray flux corresponding to this target photon field will be directly observed, with the $\nu F_{\nu}$ flux enhanced by
a factor $\delta^4$. The co-moving target photon density is therefore tightly limited by the observed hard X-ray flux (or upper 
limits thereof). In turn, a low target photon density necessitates a high co-moving density of ultrarelativistic protons in order to 
produce a given neutrino flux. This often leads to the requirement of far super-Eddington jet powers. The power requirements can be 
greatly reduced if the target photons originate outside the jet, e.g., in the Broad Line Region (e.g., \cite{Padovani19}), in a non-relativistic
sheath surrounding the relativistic inner jet (e.g., \cite{TG08}), in a radiatively inefficient accretion flow (e.g., \cite{Righi19})
or in a scenario of a possible binary or strongly bent jet structure \cite{Britzen19}. 
In this case, if the radiation field is approximately isotropic (i.e., with anisotropies on much larger angular scales than the $1/\Gamma$
angular pattern of relativistic beaming, where $\Gamma$ is the bulk Lorentz factor of the motion along the jet) in the AGN rest frame, 
then target photon energies will be boosted by a factor $\sim \delta$ into the emission-region rest frame, so the observer will see
them unboosted at energies of $E_t^{\rm obs} \sim 160 \, E_{14}^{-1} / (1 + z)$~eV, i.e., in the UV --- soft X-rays. Equally, 
the target photon energy density will be enhanced by a factor $\sim \Gamma^2$ into the emission-region rest frame, while the
observer will see the corresponding UV -- soft X-ray flux without Doppler enhancement. This allows for a much denser target photon 
field in the emission-region rest frame than in the case of a co-moving field and therefore greatly relaxes the hadronic jet power constraints. 
See, e.g., \cite{Reimer19} for the numerical evaluation of these jet power constraints in the specific case of modeling the 2014 -- 2015 neutrino 
flare from the direction of TXS 0506+056.

\subsection{\label{sec:opacity}$\gamma\gamma$ opacity constraints}

The efficient production of neutrinos through photo-pion interactions in a blazar jet inevitably leads to the production
of ultra-high-energy and VHE $\gamma$-rays through the decay of neutral pions and synchrotron emission of secondary
decay products. This has motivated systematic searches for $\gamma$-ray flaring blazars coincident with neutrino alerts
(e.g., \cite{Franckowiak20}). However, the same target photon field that leads to photo-pion production, also acts as a
source of $\gamma\gamma$ opacity for co-spatially produced $\gamma$-rays (e.g., \cite{Murase18}, \cite{Reimer19}). 
Most notably, the peak cross section (at the $\Delta^+$ resonance) for protons to interact through photo-pion production 
is about 300 times smaller than the peak $\gamma\gamma$ absorption cross section at a peak energy corresponding to 
$E_{\gamma} \sim (m_e c^2)^2 / E_t \sim 3.3 \times 10^{-5} \, E_{\nu}$. For $\gtrsim 100$~TeV neutrinos, this 
corresponds to $\sim$~GeV -- TeV $\gamma$-rays, in the energy ranges observed by {\it Fermi}-LAT and ground-based
atmospheric Cherenkov telescopes. The energy in these $\gamma\gamma$ absorbed photons will emerge from the emission
region, through electromagnetic cascading processes, in the X-rays and soft $\gamma$-rays. Consequently, a correlation between
HE and VHE $\gamma$-ray emission and neutrino activity is not necessarily expected, especially if $\gamma$-rays and neutrinos
are produced in the same emission region. On the other hand, X-ray and $\sim$~MeV $\gamma$-ray activity is expected as a
natural by-product of photo-pion processes (e.g., \cite{Murase18}, \cite{Reimer19}, \cite{Stathopoulos22}). This is also in line 
with the finding by \cite{Plavin20,Plavin21} that AGN with bright (strongly Doppler boosted) radio cores are likely sources of 
IceCube neutrinos, with radio brightness often increasing in correlation with neutrino alerts, while no systematic correlation 
with $\gamma$-ray activity is found.

\section{\label{sec:individual}Discussion of individual cases}

This section discusses those tentative associations between IceCube neutrino alerts and multi-wavelength flaring blazars which
have been studied in some detail in the literature. They are discussed in chronological order of the neutrino alerts. 

\begin{figure}[htb]
\centerline{
\includegraphics[width=5cm]{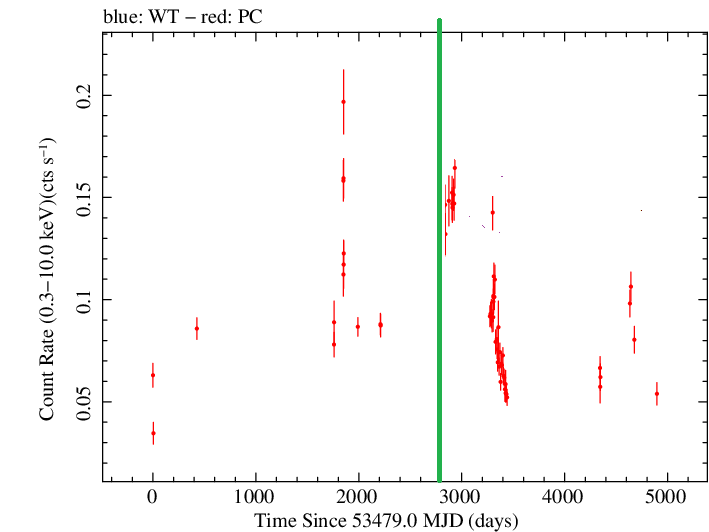} \quad \includegraphics[width=7cm]{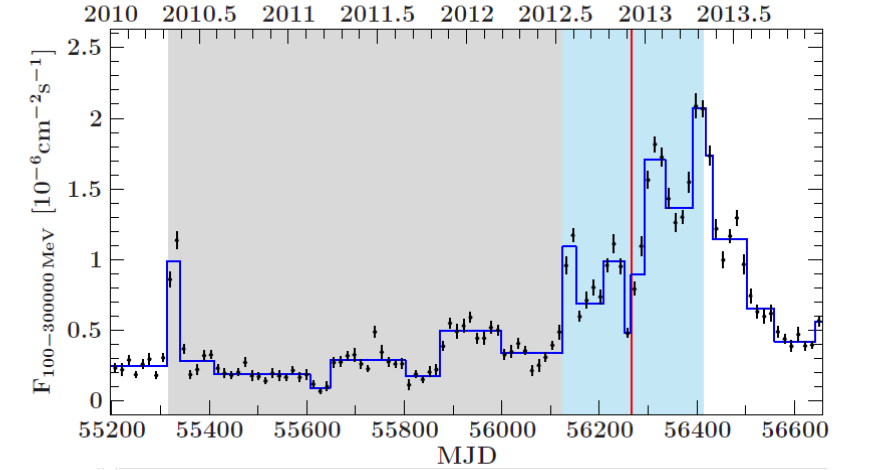}}
\caption{Swift-XRT (left --- from {\tt https://www.swift.psu.edu/monitoring/}) and Fermi-LAT (right) light curves of PKS 1424-418. 
The vertical green and red lines, respectively, indicate the arrival time of IC-35. Fermi-LAT light curve adapted from \cite{Kadler16}
--- see arXiv:1602.02012. }
\label{Fig:PKS1424_XRT_LAT}
\end{figure}

\subsection{\label{BigBird}IC-35 (``Big Bird'') and PKS 1424-418}

IC-35 was the third PeV ($E \sim 2$~PeV) neutrino detected by IceCube, on 4 December 2012. Its arrival direction, although with large
uncertainty, coincided with the FSRQ PKS 1424-418 at a redshift of $z = 1.522$. Its arrival time coincided with the rising phase of a 
multi-wavelength flare, including HE $\gamma$-rays, X-rays and radio, although Swift-XRT observations only commenced several days
after the alert (see Fig. \ref{Fig:PKS1424_XRT_LAT}). In particular, \cite{Kadler16} found that the radio outburst contemporaneous with the
neutrino alert was dominated by an increase of the flux of the radio core, possibly marking the ejection of a new jet component. 

A phenomenological, calorimetric estimate of the expected neutrino flux by \cite{Kreter20} found a very small probability of an IceCube 
neutrino detection during multi-wavelength flares of this source. More detailed physical modeling  of the SED and neutrino production
through photo-hadronic processes was performed by \cite{Gao17}. They demonstrate that neither a photo-pion dominated nor a
proton-synchrotron dominated lepto-hadronic model can simultaneously explain the SED and the PeV neutrino event. Instead, a
model with primary-electron SSC dominated $\gamma$-ray emission and only a subdominant hadronic contribution to the SED
can successfully fit the SED and predict up to 0.3 neutrino events detected by IceCube during the outburst phase.

\subsection{\label{TXS0506}IceCube-170922A and TXS 0506+056 (and PKS 0502+049?)}

The Ice-Cube detection of an $E \sim 290$~TeV neutrino on 22 September 2017, coincident with the flaring BL Lac type LSP blazar
TXS~0506+056 \cite{IceCube18a} is by far the most well studied and highest-confidence association between a neutrino event and a 
flaring blazar. What makes this association particularly compelling is the fact that an archival search for neutrino emission from the
direction of this blazar led to the identification of an excess of $\sim 13 \pm 5$ lower-energy neutrinos during a $\sim 4$~months period
around December 2014 from a direction also spatially consistent with TXS~0506+056, which, for the first time, allowed the IceCube 
Collaboration to derive a meaningful neutrino flux and spectrum (based on more than just a single neutrino) \cite{IceCube18b}. These 
associations have been discussed in a large volume of literature which has been summarized in several review papers (e.g., \cite{HK22}).  
A recent, detailed multi-epoch study of the multi-wavelength and multi-messenger emission of TXS~0506 by \cite{Petropoulou20a} also 
provides a comprehensive review of previous modeling works  with relevant references. Most works on the theoretical modeling of the 
multi-wavelength and neutrino emission agree that it is very difficult, if not impossible, to model the high-energy emission as dominated
by hadronic processes, but instead, leptonic-dominated multi-wavelength emission is favored, with only a sub-dominant hadronic 
component responsible for the detected neutrino flux, as in the case of IC-35 / PKS~1424-418 described above. Neutrino production 
is likely due to ultra-relativistic protons interacting with a UV / soft X-ray target photon field originating external to the jet (e.g., 
\cite{Keivani18,Cerruti19,Gao19,Reimer19,Petropoulou20a}). 

\begin{figure}[htb]
\centerline{
\includegraphics[width=8cm]{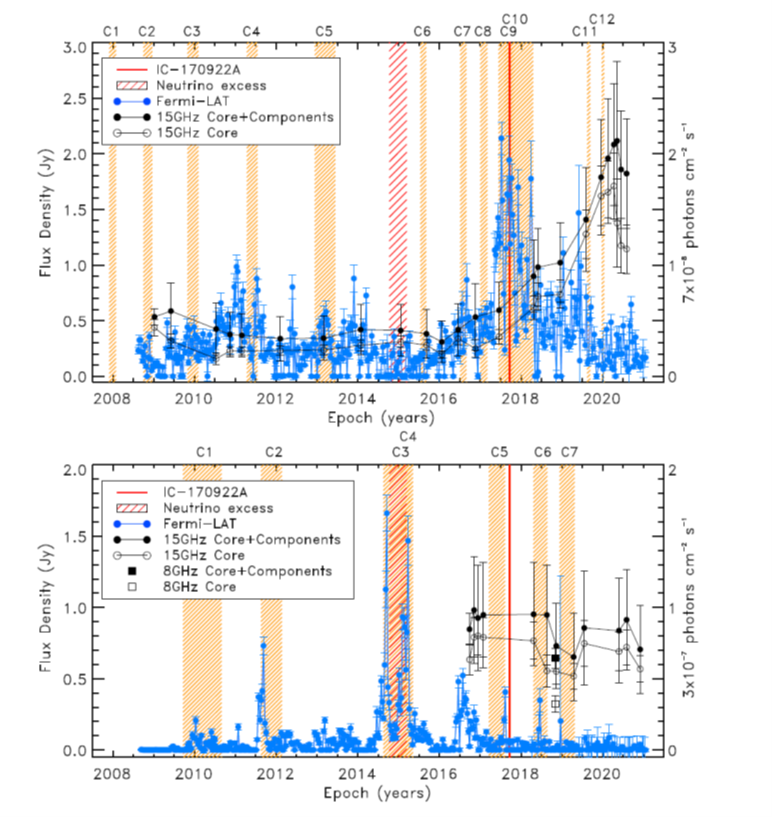}}
\caption{15 GHz radio light curves of TXS 0506+056 (top) and PKS 0502+049 (bottom). Orange vertical strips indicate the error
regions of ejection times of new jet components. The vertical solid red line indicates the arrival time of IceCube-170922A, while
the red dashed strip indicates the time of the 2014 -- 2015 neutrino flare (from \cite{Sumida22}). }
\label{Fig:TXS0506_PKS0502}
\end{figure}

A very recent finding and suggestion that has so far received little attention in the literature is based on a detailed re-analysis
of the pc-scale jet structures using archival 8 and 15 GHz very-long-baseline interferometry data, not only of TXS~0506+056, but
also of another nearby $\gamma$-ray blazar, namely PKS 0502+049 \cite{Sumida22}. Estimating the ejection times of 12 and 7 
jet components in TXS~0506+056 and PKS 0502+049, respectively, they find that the arrival time of IceCube-170922A
coincides with the estimated ejection times of two components in the jet of  TXS~0506+056, while no such coincidence was found 
for the 2014 -- 15 neutrino flare. On the contrary, the 2014 -- 2015  neutrino flare was coincident with the ejection time of a
jet component in PKS 0502+049 (see Fig. \ref{Fig:TXS0506_PKS0502}). \cite{Sumida22} thus argue that IceCube-170922A 
and the 2014 -- 15 neutrino flare might actually originate from two different blazars. However, it seems that the decomposition 
of the VLBI images into individual components as well as the identification of components across epochs appears very difficult 
in the case of these blazars, and superluminal-motion speeds up to $\gtrsim 60$~c have been found for both sources, which 
should be taken with great caution. Nevertheless, the suggestion of the two neutrino signals originating from two different 
sources should be seriously considered.

\begin{figure}[htb]
\centerline{
\includegraphics[width=8cm]{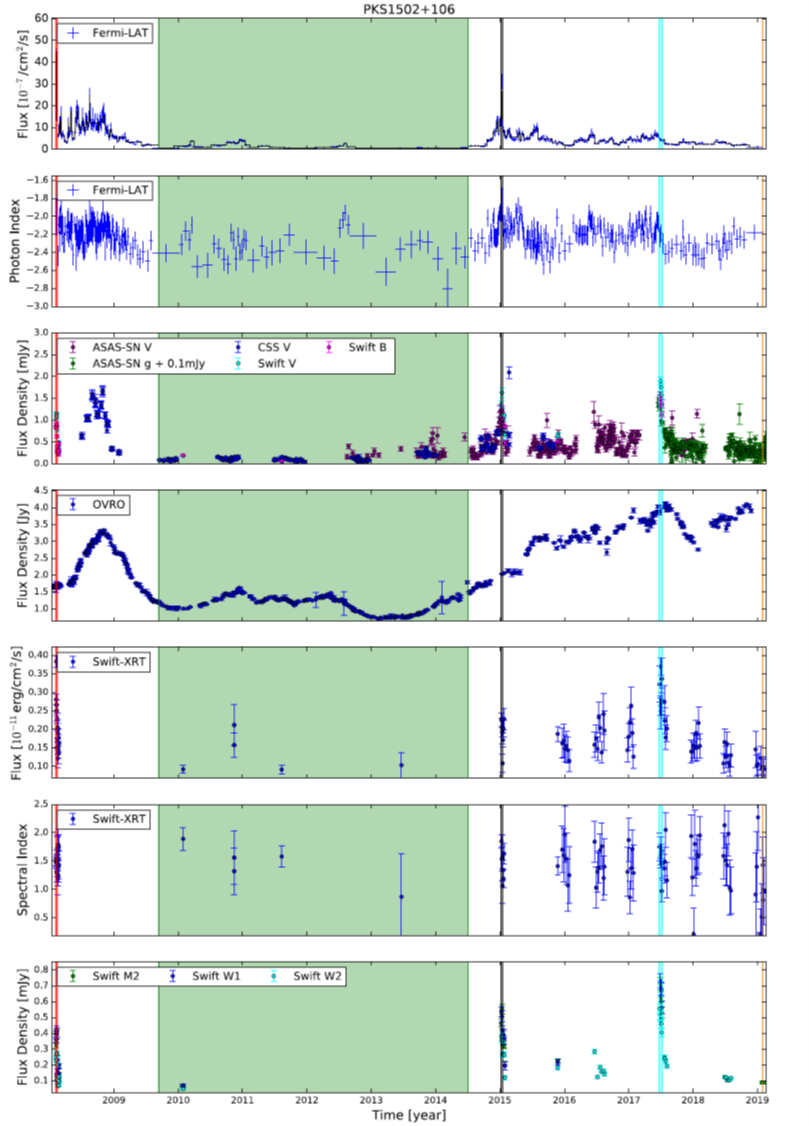}}
\caption{Multi-wavelength light curves of PKS 1502+106. The solid orange lines towards the end of the light curves indicate the
arrival time of IceCube-190730A (from \cite{Franckowiak20}).  }
\label{Fig:PKS1502_MWL_lc}
\end{figure}

\subsection{\label{PKS1502}IceCube-190730A and PKS 1502+106}

Another well-studied potential coincidence was the $\sim 300$~TeV (67~\% probability of being astrophysical) neutrino alert 
IceCube-190730A which was spatially coincident with the FSRQ PKS~1502+106 ($z = 1.84$). The coincidence attracted particular 
attention in the literature as this FSRQ is the 15$^{\rm th}$ brightest HE $\gamma$-ray source in the 4LAC \cite{4LAC}. 
This coincidence is further remarkable since the high-energy (X-ray and $\gamma$-ray) emission shows no sign of flaring
during the arrival time of IceCube-190730A, but instead, it occurs near the peak of a several-year-long radio outburst
(see Fig. \ref{Fig:PKS1502_MWL_lc}, from \cite{Franckowiak20}). This is in contrast to almost all other known neutrino --
blazar coincidences for which contemporaneous X-ray coverage exists, which consistently shows X-ray elevated states
during the time of the neutrino alert \cite{FGK22}. 

\begin{figure}[htb]
\centerline{
\includegraphics[width=12cm]{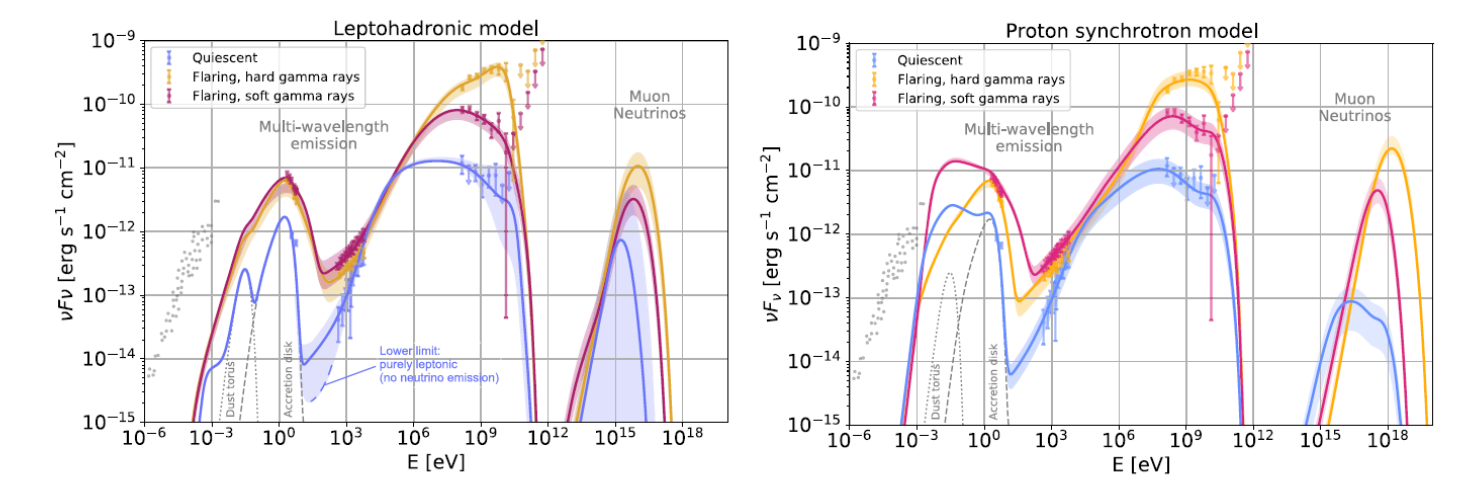}}
\caption{Model fits to the SEDs and neutrino emission from PKS 1502+106 during quiescent and flaring states, using a lepto-hadronic
(left) and a proton-synchrotron-dominated (right) model (from \cite{Rodrigues21a}).  }
\label{Fig:PKS1502_SEDfit}
\end{figure}

The multi-wavelength and neutrino emission of this source was studied in detail by \cite{Rodrigues21a} and \cite{Oikonomou21}.
\cite{Rodrigues21a} modeled the SED and neutrino emission studying various lepto-hadronic and proton-synchrotron-dominated
models and found viable model solutions --- with X-ray and $\gamma$-ray emission dominated by external Compton and 
proton-synchrotron processes, respectively (see Fig. \ref{Fig:PKS1502_SEDfit}) --- both for the quiescent and flaring states. 
They provide arguments in favor of a lepto-hadronic scenario, in particular a lower power demand, but such a scenario also 
predicts detectable neutrino emission in flaring states (in particular, prior to 2010), which should be found in archival searches. 

In contrast, \cite{Oikonomou21} investigated various scenarios with leptonic (SSC and external-Compton) scenarios for the
high-energy emission from PKS 1502+106 with only a sub-dominant hadronic component responsible for neutrino production
(more in line with the majority of findings for TXS~0506+056). They found that both an SSC and an EC scenario are viable and 
may predict $\sim 1$ muon neutrino during the life time of IceCube, with power requirements signficantly below the source's 
Eddington limit, while at the same time being consistent with blazars powering the observed ultra-high-energy cosmic-ray flux.

\subsection{\label{200107}IceCube-200107 and 3HSP~J095507.9+35510}

IceCube-200107 was an IceCube High-Energy Starting Event (HESE) with highly uncertain energy ($E_{\nu} \sim 330^{+2230}_{-270}$~TeV).
It was found to be spatially coincident with the high-synchrotron-peaked BL Lac object 3HSP J095507.9+35510 (= BZB J0955+3551) 
\cite{Giommi20a} at a redshift of $z = 0.5573$ 
\cite{Paiano20,Paliya20}. {\it Swift}-XRT ToO observations one day and NuSTAR ToO observations 4 days after the neutrino alert found the blazar 
in a bright X-ray flare, but exhibiting only moderate $\gamma$-ray activity \cite{Giommi20a,Paliya20}. Incidentally, there was one other 
$\gamma$-ray blazar within the error region of the neutrino event, namely 4FGL~J0957.8+3423, which, however, did not exhibit any significant 
flux enhancement in any waveband \cite{Krauss20}. 3HSP J~095507.9+35510 is therefore considered the more likely counterpart of IceCube-200107. 

\begin{figure}[htb]
\centerline{
\includegraphics[width=6.5cm]{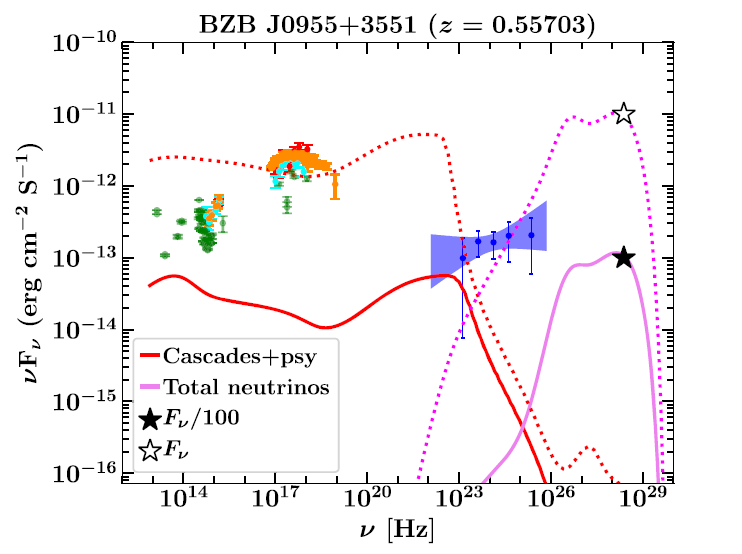}
\quad
\includegraphics[width=6cm]{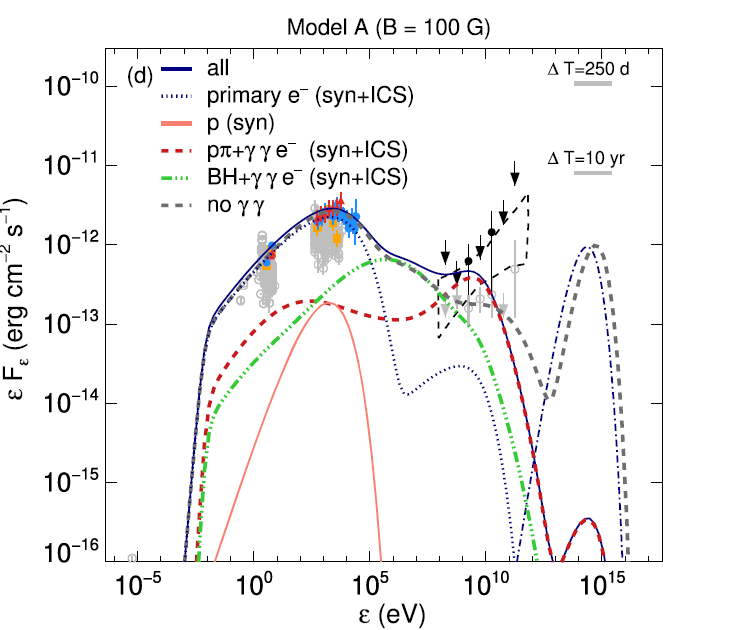}}
\caption{{\it Left:} Model calculations of the neutrino flux and resulting p$\gamma$-induced cascade emission assuming a neutrino
flux corresponding to 1 neutrino in the life-time of the current neutrino alert system (dashed) and accounting for the Eddington bias
by reducing the flux by a factor 100 (solid), respectively (from \cite{Paliya20}). {\it Right:} Example of a lepto-hadronic one-zone 
model SED fit + neutrino expectations from \cite{Petropoulou20b}.  }
\label{Fig:200107fits}
\end{figure}

Estimates of requirements on jet power in relativistic protons and on the target photon field in BZB J0955+3551 to produce a neutrino
flux corresponding to an association with IceCube-200107, along with numerical simulations, were presented in \cite{Paliya20}. They 
found that the resulting p$\gamma$-induced cascade emission would significantly overshoot the observed optical flux from the source
if the neutrino flux estimate is based on the occurrence of 1 HESE during the lifetime of the alert system (see Fig. \ref{Fig:200107fits}, 
left panel). However, the actual neutrino flux may be significantly lower than this estimate since there are likely many blazars with similar 
properties, also exhibiting a similar neutrino flux, from which no neutrinos have been detected. This is known as the Eddington bias 
\cite{Strotjohann19} and motivates a lowering of the neutrino flux estimate by a factor of $\sim 100$ (i.e., assuming a probability of 
$\sim 1$~\% of detecting a HESE muon neutrino from this particular source). This Eddington-bias-corrected neutrino flux can be
produced without violating any observational constraints and with moderate proton power requirements of $\lesssim 10^{47}$~erg~s$^{-1}$.
While a target photon field originating external to the jet is preferred, a co-moving target photon field can not be ruled out, contrary 
to the case of TXS~0506+056. Also, due to the required dense target photon field, $\gamma\gamma$ absorption suppresses any
co-spatially produced HE $\gamma$-ray emission, implying that no neutrino -- $\gamma$-ray correlation is necessarily expected,
as discussed above in \S \ref{sec:opacity} \cite{Paliya20}.

Various scenarios for the production of the multi-wavelength emission of 3HSP~J095507.9+35510 and the IceCube-200107 event
were explored in detail by \cite{Petropoulou20b}. They investigate one-zone lepto-hadronic models, and also consider alternative
scenarios, such as an intergalactic cascade scenario based on a high-energy cosmic-ray beam originating from the blazar jet,
interacting with the intergalactic medium (see Fig. \ref{Fig:200107fits}, right panel, for an example of a lepto-hadronic model fit). 
In agreement with the Eddington-bias argument mentioned above, they find a probability of $\sim 1$ -- a few~\% of detecting
1 muon neutrino in 10 years from 3HSP~J095507.9+35510 for the most optimistic lepto-hadronic models, in which they also
predict strong internal $\gamma\gamma$ absorption and, hence, a suppression of HE and VHE $\gamma$-ray emission from
the neutrino production region. The same issue was encountered by \cite{Gasparyan22}, who investigated both a hadronic
(proton-synchrotron dominated) and a lepto-hadronic (SSC dominated $\gamma$-ray emission) scenario for this source.
While moderate power requirements (of the order of the Eddington luminosity) were found for their hadronic model, their
lepto-hadronic model required proton kinetic jet powers of $\sim 10^{51}$~erg~s$^{-1}$ which seem unrealistic.

\subsection{\label{211125}IceCube-211125 and AT2021afpi / 4FGL J0258.1+2030}

Two potential counterparts were identified for the IceCube-211125 neutrino, with an estimated energy of $E \sim 117$~TeV (39~\% 
probability of being astrophysical). One is the young, narrow-line, He-rich nova AT2021afpi \cite{Stein21,Taguchi21}, which was 
detected as an optical transient by the MASTER network of robotic telescopes \cite{Zhirkov21a}. It was found to be in an
X-ray outburst around the time of the IceCube-211125 alert \cite{Paliya21}. This would be the first such association of a classical nova
outburst with a neutrino event. 

The second potential counterpart is the Fermi-LAT source 4FGL~J0258.1 +2030, a blazar of unknown type with uncertain
redshift ($z \sim 2.2$). It is consistent with a compact radio source in a high state, detected by the TELAMON program,
which is also spatially coincident with another IceCube Bronze neutrino event, namely IceCube-191231A \cite{Kadler21}. 
If this association is true, it would be the first ever source of more than one single $> 100$~TeV neutrino, in which 
case the Eddington-bias argument would no longer apply. Unfortunately, there is very little information about this source 
at other wavelengths, and, to our knowledge, no theoretical modeling of this source and the potentially associated neutrino 
emission has been published.

\begin{figure}[htb]
\centerline{
\includegraphics[width=12cm]{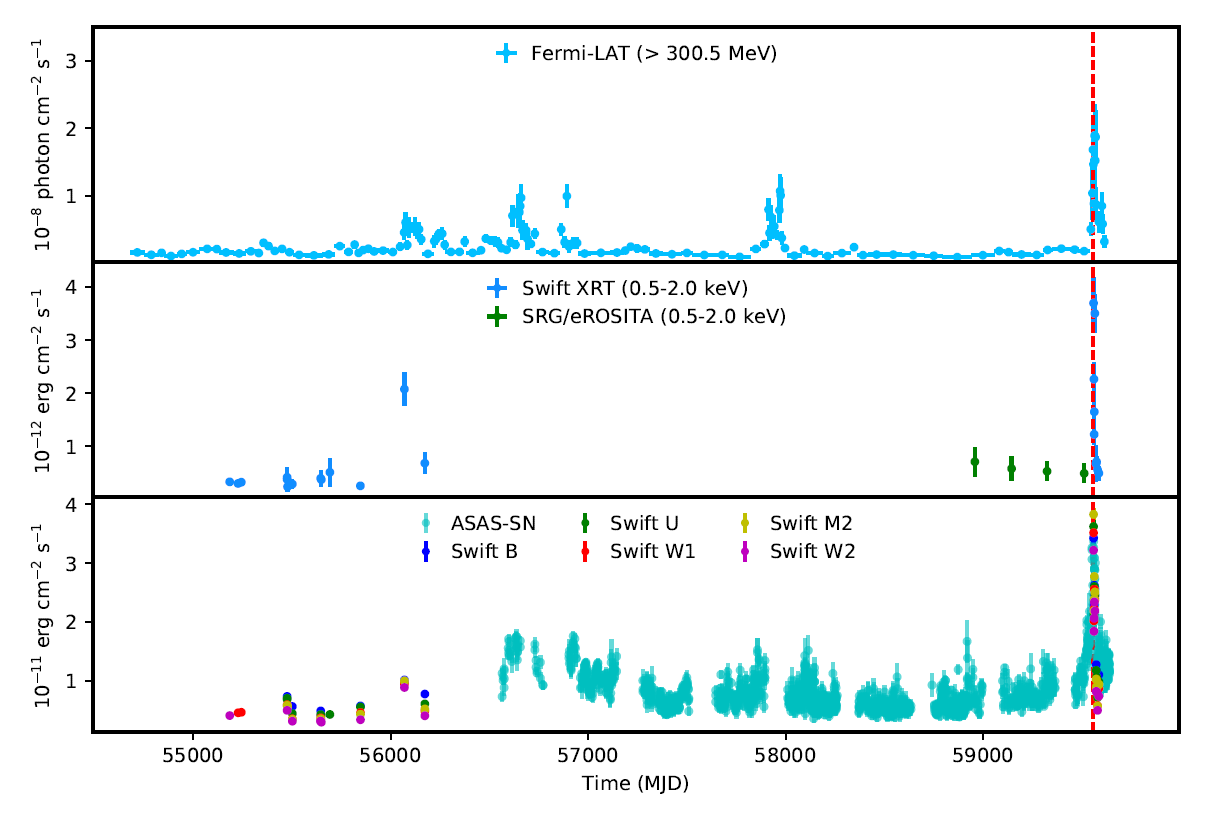}}
\caption{{\it Fermi}-LAT (top), X-ray (middle), and optical -- UV (bottom) light curves of PKS 0735+17 (from \cite{Sahakyan22}). 
The red dashed vertical line indicates the arrival time of IceCube-211208.  }
\label{Fig:PKS0735_lc}
\end{figure}

\subsection{\label{211208}IceCube-211208 and PKS~0735+17}

IceCube-211208 was a track-like event with an estimated energy of $E \sim 172$~TeV with a relatively large statistical 
90~\% localization error region of $\sim 13$ square degrees. The BL Lac type ISP blazar PKS~0735+17 ($z = 0.45$) is 
actually located slightly outside this error region, but was in a historical outburst in HE $\gamma$-rays ({\it Fermi}-LAT), X-rays
({\it Swift}-XRT), and optical -- UV at the time of the neutrino alert (see Fig. \ref{Fig:PKS0735_lc}, from \cite{Sahakyan22}). In 
addition, the source is positionally consistent with the (rather large) localization error regions of a Baikal-GVD detected neutrino
event ($E \sim 43$~TeV) 3.95 hours after the IceCube-211208 event \cite{Dzhilkibaev21}, of a GeV neutrino detected 
4 days earlier by the Baksan Underground Scintillation Telescope \cite{Petkov21}, as well as an $\sim 18$~TeV neutrino
a week after IceCube-211208, found in an archival search of data from the Km3NeT undersea neutrino detectors, currently 
under construction in the Meditarranean Sea \cite{Filippini22}. It is this multitude of potential neutrino associations as well
as the exceptional flaring state of PKS~0736+17 which drew the attention of the community to this source as a possible,
prolific VHE neutrino source. 

\begin{figure}[htb]
\centerline{
\includegraphics[width=12cm]{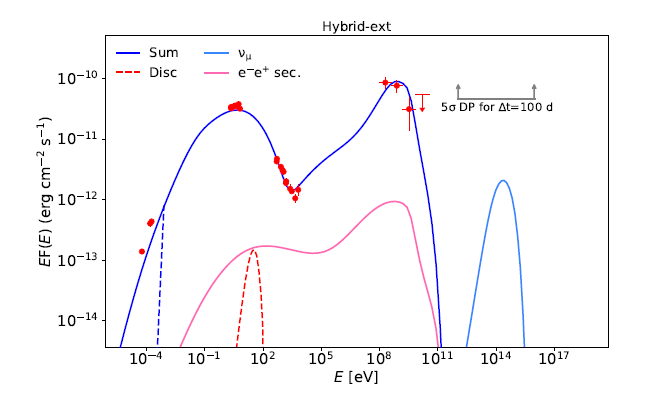}}
\caption{SED of PKS~0735+17, along with the most favorable lepto-hadronic single-zone model fit, including a BLR 
radiation field as the target photon field for p$\gamma$ processes, from \cite{Sahakyan22}.  }
\label{Fig:PKS0735_BLR}
\end{figure}

\cite{Sahakyan22} investigated several lepto-hadronic single-zone emission scenarios for the multi-wavelength and
neutrino emission from PKS~0735+178. Their most optimistic scenario requires a jet with a kinetic luminosity close to 
the Eddington limit of the source, where relativistic protons interact with an external radiation field from the BLR (see 
Fig. \ref{Fig:PKS0735_BLR}), resulting in a 6.7~\% probability of detecting one muon / antimuon neutrino during a
3-week flaring period.

\section{\label{sec:summary}Summary and discussion}

We have provided an overview of the physical conditions required in blazar jets to produce astrophysical VHE neutrinos 
detected by IceCube. We have demonstrated that protons need to be accelerated to $\gtrsim$~PeV energies, and a
target photon field of UV / X-ray energies in the co-moving frame of the neutrino production regions is required. On
energetics grounds, photon fields originating external to the jet are generally favored. Due to the dense target photon
field required for efficient neutrino production, the emission region is likely to be opaque to internal $\gamma\gamma$
absorption. Therefore, a correlation between neutrino production and $\gamma$-ray flaring activity is not necessarily
expected. However, due to cascading of the internally absorbed HE and VHE $\gamma$-ray emission, enhanced X-ray
-- soft $\gamma$-ray is a natural byproduct of p$\gamma$-induced neutrino production. Another pattern that seems
to emerge is the coincidence of radio brightening and the ejection of new jet components with neutrino emission,
indicating that neutrino production is likely occurring within the central $\sim$~pc of the jet. 

After the first relatively-high-confidence association between IceCube-170922A and TXS~0506+056, the number of
potential associations between flaring blazars and IceCube neutrino alerts has been steadily increasing. We have
provided a brief overview of the best-known associations and, where available, results of model calculations to reproduce
coincident multi-wavelength and neutrino emission. In most cases, the detailed modeling results confirm the general
expectations outlined in Section \ref{sec:energetics}. In particular, it is often found that internal $\gamma\gamma$
absorption suppresses the co-spatially produced HE and VHE $\gamma$-ray emission and that neutrino production
is accompanied by high states in X-rays and soft $\gamma$-rays. Also, in many cases, p$\gamma$ interactions with
external radiation fields are indeed favored over co-moving (primary electron synchrotron) target photon fields. 

A particularly exciting development is the possibility that at least one blazar (4FGL~J0258.1+2030) could be associated
with two IceCube neutrino events and that PKS 0735+17 could be the counterpart not only to one IceCube event, but
also to neutrino events detected by Baikal-GVD, Baksan, and KM3NeT. If such multiple associations are confirmed by
future observations, it would allow for a more meaningful determination of VHE neutrino fluxes from such sources,
and it would invalidate the Eddington-bias argument that models can be considered successful if they predict only a
probability of a few \% to detect a neutrino from a single source, accounting for the fact that there are many other
sources with similar properties which may exhibit the same neutrino flux.

\noindent
{\bf Acknowledgments}

\noindent
The work of MB is supported by the South African Research Chairs Initiative (SARChI) of the Department of Science and Innovation and the National Research Foundation\footnote{Any opinion, finding and conclusion or recommendation expressed in this material is that of the authors, and the NRF does not accept any liability in this regard.} of South Africa through SARChI grant UID 64789.

We thank Matthias Kadler, Viktor Sumida, Anna Franckowiak, Xavier Rodrigues, Maria Petropoulou, and Narek Sahakyan for granting us
permission to use figures from their publications in this review, and for helpful comments on the manuscript.

\end{document}